\documentclass[preprint,preprintnumbers,amsmath,amssymb]{revtex4}
\usepackage{graphics}
\usepackage{dcolumn}
\usepackage{bm}

\begin{document}

\title{Integral transformation solution of free-space cylindrical vector beams and
prediction of modified-Bessel-Gaussian vector beams}

\author{Chun-Fang Li}

\affiliation{Department of Physics, Shanghai University, Shanghai 200444, P. R. China}

\affiliation{State Key Laboratory of Transient Optics and Photonics, Xi'an Institute of Optics and
Precision Mechanics of CAS, Xi'an 710119, P. R. China}

\date{\today}

\begin{abstract}

A unified description of the free-space cylindrical vector beams is presented, which is
an integral transformation solution to the vector Helmholtz equation and the
transversality condition. The amplitude 2-form of the angular spectrum involved in this
solution can be arbitrarily chosen. When one of the two elements is zero, we arrive at
either transverse-electric or transverse-magnetic beam mode. In the paraxial condition,
this solution not only includes the known $J_1$ Bessel-Gaussian vector beam and the
axisymmetric Laguerre-Gaussian vector beam that were obtained by solving the paraxial
wave equations, but also predicts two new kinds of vector beam, called the
modified-Bessel-Gaussian vector beam.

\end{abstract}

\maketitle


A free-space cylindrical vector beam is of spatially inhomogeneous polarization that is
rotationally symmetric with respect to the propagation axis. Due to this unique
polarization characteristic, the cylindrical vector beam has attracted much attention in
both optical physics \cite{Greene, Youngworth, Dorn, Guo} and applied optics
\cite{Ashkin, Zhan, Biss}. The mathematical expression for a paraxial cylindrical vector
beam is usually obtained by solving the paraxial wave equation \cite{Davis, Jordan, Hall,
Tovar-C, Tovar, Bandres}. And the known kinds of cylindrical vector beam include the
Bessel-Gaussian \cite{Hall} and Laguerre-Gaussian beams \cite{Tovar}. In this paper, we
present a unified description of the cylindrical vector beam, from which we obtain for
the first time two new kinds of vector beam, called the modified-Bessel-Gaussian vector
beam. This beam description represents an integral transformation solution to the vector
Helmholtz equation and the transversality condition.

Consider a light beam that propagates in the positive $x$ direction in source-free space.
The electric-field vector $\mathbf{E}$ of the beam satisfies the vector Helmholtz
equation,
\begin{equation} \label{Helmholtz equation}
\nabla^2 \mathbf{E}(\mathbf{x})+k^2 \mathbf{E}(\mathbf{x})=0,
\end{equation}
subject to the transversality condition
\begin{equation} \label{transverse condition}
\nabla \cdot \mathbf{E}(\mathbf{x})=0.
\end{equation}
For a rotationally symmetric beam with respect to its propagation axis, it is convenient
to make use of the cylindrical coordinate system, in which $\mathbf{x}= x \mathbf{e}_x+
\mathbf{r}$, where $\mathbf{r}= r \mathbf{e}_r= \mathbf{e}_y r \cos \phi+ \mathbf{e}_z r
\sin \phi$ is the polar coordinate. It has been shown that the electric-field vector of
the beam can be represented by the following integral over the plane-wave angular
spectrum \cite{Li},
\begin{equation}\label{electric field of a beam}
\mathbf{E}(\mathbf{x})= \frac{1}{2 \pi} \int \int_{k_{\rho}<k}
\mathbf{A}(k_{\rho},\varphi) \exp(i \mathbf{k} \cdot \mathbf{x}) d \Omega,
\end{equation}
where $\mathbf{k}= k_x \mathbf{e}_x+ \mathbf{k}_{\rho}$ is the wavevector of the plane
wave, $\mathbf{k}_{\rho}= k_{\rho} \mathbf{e}_{\rho}= \mathbf{e}_y k_{\rho} \cos \varphi+
\mathbf{e}_z k_{\rho} \sin \varphi$, $k_x= (k^2-k^2_{\rho})^{1/2}$,
\begin{equation} \label{vector in 2-form}
\mathbf{A} (k_{\rho},\varphi)= \mathrm{P} \tilde{A} (k_{\rho},\varphi)
\end{equation}
is the amplitude vector of the angular spectrum,
\begin{equation} \label{two-form amplitude}
\tilde{A} (k_{\rho},\varphi)= \left(
                                    \begin{array}{c}
                                       A_s\\A_p
                                    \end{array}
                              \right)
\end{equation}
is the amplitude two-form \cite{Li} of the angular spectrum,
\begin{equation} \label{extension matrix}
\mathrm{P}=\left(
                 \begin{array}{cc}
                     s_x & p_x\\s_y & p_y\\s_z & p_z
                 \end{array}
           \right)
          \equiv (\begin{array}{cc} \mathbf{s} & \mathbf{p} \end{array})
\end{equation}
is a $3 \times 2$ matrix that plays the role of extending the amplitude two-form $\tilde{A}$ to
the three-component amplitude vector $\mathbf{A}$ and is thus referred to as the extension matrix,
$\mathbf{s}=\left( \begin{array}{c} s_x\\s_y\\s_z \end{array} \right)$ and $\mathbf{p}=\left(
\begin{array}{c} p_x\\p_y\\p_z \end{array} \right)$ are unit vectors that embody the vectorial
nature of the beam and are given by \cite{Li, Herrero}
\begin{equation}
\label{unit s} \mathbf{s}= \mathbf{e}_{\varphi}
\end{equation}
and
\begin{equation}
\label{unit p} \mathbf{p}= -\frac{k_{\rho}}{k} \mathbf{e}_x+ \frac{k_x}{k}
\mathbf{e}_{\rho},
\end{equation}
respectively. Because $\mathbf{s}$, $\mathbf{p}$, and $\mathbf{k}$ are mutually
orthogonal, Eq. (\ref{electric field of a beam}), together with Eqs. (\ref{vector in
2-form})-(\ref{unit p}), constitutes an integral transformation solution to the wave
equations (\ref{Helmholtz equation}) and (\ref{transverse condition}). It deserves
mentioning that the amplitude two-form $\tilde{A}$ of the angular spectrum can be
arbitrarily chosen.

Eqs. (\ref{unit s}) and (\ref{unit p}) show that only element $p_x$ of the extension
matrix produces the longitudinal component. If $\tilde{A}=\left(
\begin{array}{c} A_s\\0
\end{array} \right)$, we arrive at the transverse-electric beam mode \cite{Davis}. The
principle of duality predicts that $\tilde{A}=\left( \begin{array}{c} 0\\A_p
\end{array} \right)$ corresponds to the transverse-magnetic beam mode.

In this paper, we consider only the following amplitude two-form that is independent of
the azimuthal angle $\varphi$,
\begin{equation}\label{2-form amplitude}
\tilde{A}= \left( \begin{array}{c} l_s \\ l_p \end{array} \right) A(k_\rho)
         \equiv \tilde{l} A(k_\rho),
\end{equation}
where $l_s$ and $l_p$ are constants, $\tilde{l}= \left( \begin{array}{c} l_s \\ l_p
\end{array} \right)$ describes the polarization state of the angular spectrum and is
assumed to satisfy the normalization condition $|l_s|^2+|l_p|^2=1$, and $A(k_{\rho})$ is
the amplitude distribution of the angular spectrum. Let us consider the following
Gaussian-like distribution function,
\begin{equation}\label{angular distribution}
A(k_{\rho})= A_0 \exp \left( -\frac{w^2_0}{2} k^2_{\rho} \right) A_m(k_{\rho}),
\end{equation}
where $A_0$ is a constant and $w_0$ is the characteristic width in the transverse
dimension. The Gaussian factor guarantees that the beam carries finite energy. Different
choices of the modulation factor, $A_m$, will correspond to different kinds of beam as
will be shown below.

For the sake of simplicity, we discuss only the paraxial beam, for which condition
\begin{equation} \label{paraxial condition}
\Delta \theta= \frac{1}{k w_0} \ll 1
\end{equation}
holds \cite{Lax}, where $\Delta \theta$ is half the divergence angle that is determined
by the Gaussian factor in Eq. (\ref{angular distribution}). Under this paraxial
condition, Eq. (\ref{electric field of a beam}) can be rewritten as
\begin{equation}\label{electric field of paraxial beam}
\mathbf{E}(\mathbf{x})= \frac{1}{2\pi} \int_{-\infty}^{\infty} k_{\rho} dk_{\rho} \int_{0}^{2
\pi} d\varphi \mathbf{A} (k_{\rho},\varphi) \exp(i \mathbf{k} \cdot \mathbf{x}),
\end{equation}
where the integration limits have been extended to $\pm \infty$ for the variable
$k_{\rho}$. And the Gaussian factor in Eq. (\ref{angular distribution}) indicates that
the quantity $\frac{k_{\rho}}{k}$ in the extension matrix can be regarded as a small
number in comparison with unity when integral (\ref{electric field of paraxial beam}) is
considered. This explains why the longitudinal component of a Gaussian-like paraxial beam
is of the first order in comparison with the zeroth-order transverse component
\cite{Lax}. Substituting Eqs. (\ref{vector in 2-form}) and (\ref{extension
matrix})-(\ref{angular distribution}) into Eq. (\ref{electric field of paraxial beam})
and with the help of the following expansion,
\begin{equation} \label{expansion}
\exp(i \rho \cos \psi)= \sum_{m=-\infty}^{\infty} i^m J_m (\rho) \exp(im \psi),
\end{equation}
where $J_m$'s  are the Bessel functions of the first kind, we obtain for the
electric-field vector,
\begin{equation} \label{electric field}
\mathbf{E} (\mathbf{x})= [i(l_s \mathbf{e}_{\phi}+ l_p \mathbf{e}_r) E_T (r,x)- l_p
\mathbf{e}_x E_L (r,x)] \exp(ikx),
\end{equation}
where
\begin{equation} \label{transverse component}
E_T (r,x)= \int_0^{\infty} A'(k_{\rho}) J_1 (r k_{\rho}) k_{\rho} dk_{\rho},
\end{equation}
\begin{equation} \label{longitudinal component}
E_L (r,x)= \int_0^{\infty} \frac{k_{\rho}}{k} A'(k_{\rho}) J_0 (r k_{\rho}) k_{\rho}
dk_{\rho},
\end{equation}
$$
A'(k_{\rho})= A_0 \exp \left( -\frac{w^2}{2} k^2_{\rho} \right) A_m(k_{\rho}),
$$
$$
w^2= w^2_0 \left( 1+i \frac{x}{x_R} \right),
$$
and $x_R=k w^2_0$, which represents the diffraction length. In deriving Eq.
(\ref{electric field}), we have also made (i) the paraxial approximation \cite{Enderlein}
$k_x \approx k- \frac{k^2_{\rho}}{2k}$ in the exponential factor $\exp(i \mathbf{k} \cdot
\mathbf{x})$, (ii) and the zeroth-order approximation $\mathbf{p}_{\rho}= \frac{k_x}{k}
\mathbf{e}_{\rho} \approx \mathbf{e}_{\rho}$ in the extension matrix.

Eq. (\ref{electric field}) describes bound beams that are axisymmetric with respect to
the propagation axis not only in the polarization but also in the complex amplitude. By
``axisymmetric" we mean ``invariant" under arbitrary rotation about the axis. The first
term on the right side is the transverse component. Its amplitude, given by Eq.
(\ref{transverse component}), is the Hankel transformation \cite{Andrews-AR} of order one
of the function $A'$ and is of the zeroth order. The second term is the longitudinal
component. Its amplitude, given by Eq. (\ref{longitudinal component}), is the Hankel
transformation of order zero of the function $\frac{k_{\rho}}{k}A'$ and is therefore of
the first order, $\sim \frac{k_{\rho}}{k}$. So the longitudinal component is much smaller
than the transverse component \cite{Lax}. Neglecting the small longitudinal component,
the beam is dark on the axis $r=0$ and is locally polarized elliptically with the same
polarization state as that of the angular spectrum, $\tilde{l}$.

Let us now look at a few examples, paying our attention mainly to the amplitude of the
transverse component.

\textit{1 Doughnut modified-Bessel-Gaussian vector beams I} For the simplest modification
factor,
$$
A_m=1,
$$
we obtain for the amplitude of the transverse component,
\begin{eqnarray}
E_T (r,x) & = & \frac{\sqrt{2 \pi}}{4 w^3} A_0 r \exp\left( -\frac{r^2}{4 w^2} \right) \times \nonumber \\
          &   & \left[ I_0 \left( \frac{r^2}{4 w^2} \right)- I_1 \left( \frac{r^2}{4 w^2} \right) \right], \nonumber
\end{eqnarray}
where $I_0$ and $I_1$ are the modified Bessel functions of the first kind. Due to the
linear factor $r$, the beam is dark on the axis. In addition, there is only one bright
ring in the transverse intensity distribution. So this is a doughnut beam.

\begin{figure}[ht]
\includegraphics{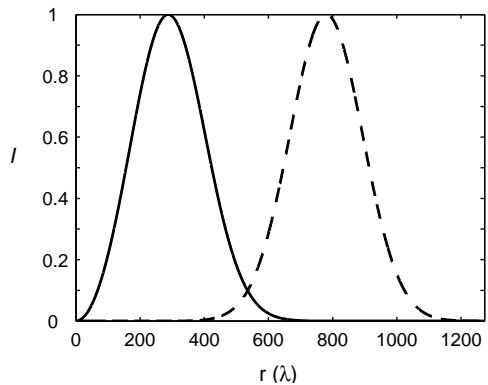} \caption{Dependence of normalized transverse intensity on
the radial coordinate at the focal plane, where $k w_0=1000$ and $r$ is in units of wavelength
$\lambda$. Solid: $\beta=0.002k$; dashed: $\beta=0.005k$.} \label{radius}
\end{figure}

\textit{2 Doughnut modified-Bessel-Gaussian vector beams II} If we choose for the
modification factor,
$$
A_m= \exp\left( -\frac{w^2_0}{2} \beta^2 \right) J_1 (\beta w^2_0 k_{\rho}),
$$
where $\beta$ is a constant, we have
\begin{eqnarray}
E_T (r,x) & = & \frac{A_0}{w^2} \exp\left[ -\frac{\beta^2 w^2_0}{2} \left( 1+
\frac{w^2_0}{w^2} \right) \right] \times \nonumber \\
          &   & \exp\left(-\frac{r^2}{2 w^2}\right) I_1 (\frac{w^2_0}{w^2} \beta r). \nonumber
\end{eqnarray}
This is also a doughnut beam. The radius of the bright ring expands with the increase of
the value of $\beta$. But the width of the ring changes little. In Fig. \ref{radius} is
shown the dependence of the transverse intensity, $I=|E_T|^2$, on the radial coordinate
$r$ at the focal plane $x=0$, where $k w_0=1000$, the intensity is normalized to unity,
$r$ is in units of wavelength $\lambda$, the solid curve is for $\beta=0.002k$, and the
dashed curve is for $\beta=0.005k$.

To the best of our know, this is the first time to observe theoretically the
above-mentioned 2 kinds of vector beam. Since the modified-Bessel-Gaussian scalar beams
show an elongated diffraction-free region \cite{Ruschin} in comparison with the
fundamental Gaussian beam, the propagation properties of the modified-Bessel-Gaussian
vector beams deserve investigation in detail. This is beyond the scope of this paper and
will be presented elsewhere.

\textit{3 Bessel-Gaussian vector beams} With a modification factor containing the
modified Bessel function of the first kind of order one,
$$
A_m= \exp\left( -\frac{w^2_0}{2} \beta^2 \right) I_1 (\beta w^2_0 k_{\rho}),
$$
we find
$$
E_T (r,x)= \frac{A_0}{w^2} J_1 (\frac{w^2_0}{w^2} \beta r) \exp\left[ -\frac{1}{2 w^2}
                   (r^2+ i \beta^2 w^2_0 \frac{x}{k}) \right].
$$
If $\tilde{l}=\left( \begin{array}{c} 1\\0 \end{array} \right)$, we arrive at the
azimuthally polarized $J_1$ Bessel-Gaussian vector beam, the same as was obtained by
solving the paraxial wave equation \cite{Jordan}.

\textit{4 Laguerre-Gaussian vector beams} Furthermore, with a modification factor
containing the associated Laguerre polynomial,
$$
A_m= \frac{k_{\rho}}{k} L^1_n (-\frac{\alpha^2}{2} k^2_{\rho}),
$$
where $\alpha$ is a constant, we have
\begin{eqnarray}
E_T (r,x) & = & \frac{A_0}{k w^4} \left( 1+ \frac{\alpha^2}{w^2} \right)^n r \times \nonumber \\
          &   & \exp\left( -\frac{r^2}{2 w^2} \right)
                L^1_n \left( -\frac{\alpha^2}{w^2+\alpha^2} \frac{r^2}{2 w^2} \right). \nonumber
\end{eqnarray}
This kind of vector beam includes the axisymmetric Laguerre-Gaussian vector beam
discussed in Ref. \cite{Tovar}.

When the amplitude two-form $\tilde{A}$ depends on the azimuthal variable $\varphi$ in
the wavevector space, the vector beam produced by Eq. (\ref{electric field of a beam})
will in general be no longer axisymmetric. Suitable choices will lead to cylindrical
vector beams of topological charges \cite{Stalder, Hall, Tovar}.

This work was supported in part by the National Natural Science Foundation of China
(Grant 60377025), Science and Technology Commission of Shanghai Municipal (Grant
04JC14036), and the Shanghai Leading Academic Discipline Program (T0104).

\end{document}